\documentclass[aps,prb,twocolumn,amsmath,showpacs,amssymb]{revtex4}
\usepackage{amsmath}
\usepackage{graphicx}
\usepackage{amssymb}
\newcommand{\nn}{\nonumber}

\begin{document}

\title{Quantum criticality with multiple dynamics}

\author{Tobias Meng}
\author{Achim Rosch}
\author{Markus Garst}
\affiliation{Institut f\"ur Theoretische Physik, Universit\"at zu K\"oln,
Z\"ulpicher Str.~77, 50937 K\"oln, Germany}

\begin{abstract}
Quantum critical systems with multiple dynamics possess not only one but several time scales, $\tau_i \sim \xi^{z_i}$, which diverge with the correlation length $\xi$. We investigate how scaling predictions are modified for the simplest case of multiple dynamics characterized by two dynamical critical exponents, $z_>$ and $z_<$. We argue that one should distinguish 
the case of {\it coupled} and {\it decoupled} multiple dynamic scaling depending on whether there exists a scaling exponent 
which depends on both $z_i$ or not. 
As an example, we study generalized $\Phi^4$-theories with multiple dynamics below their upper critical dimension, $d+z_<<4$. We identify under which condition coupled scaling is generated.  In this case the interaction of quantum and classical fluctuations leads to an emergent dynamical exponent, $z_e=\frac{z_>}{\nu (z_>-z_<)+1}$.
\end{abstract}

\date{\today}

\pacs{71.10.Hf, 71.27.+a, 73.43.Nq}
\maketitle

Quantum criticality has become a ubiquitous theme in condensed matter physics. The anomalous thermodynamics and transport properties close to a quantum critical point are fascinating as they defy paradigmatic theories of normal metals and insulators, and they are now frequently invoked to explain the unusual behavior observed in a wide class of systems ranging from heavy-fermion compounds, organic materials and high-$T_c$ superconductors to ultracold atoms.

Quantum critical points are not only characterized by the symmetries of the system and the critical degrees of freedom, but also by their specific dynamics \cite{SachdevBook,Loehneysen07}. However, often these attributes are not easily identified and simplifying assumptions are applied. For example, the quantum critical scaling hypothesis, that is commonly employed for the interpretation of experiments, assumes 
the critical behavior to be dominated by a single time scale $\tau$ that diverges together with the correlation length, $\xi$, according to $\tau \sim \xi^z$, where $z$ is the dynamical exponent. At finite temperatures $T$, scaling then implies the presence of a thermal length scale, $\xi_T \sim T^{-1/z}$, characterizing the distance beyond which thermal dominate over quantum fluctuations.
However, often various critical degrees of freedom with their respective dynamics coexist giving rise to multiple thermal lengths. In particular, in quantum critical metals critical collective modes interact with critical fermionic quasi-particles that are both governed by different dynamics. In such cases, scaling assumptions become ambiguous as certain physical observables could be determined by one or the other dynamics of the different critical degrees of freedom \cite{Gegenwart07,Belitz01,Metlitski10,Loehneysen07,Zacharias09,Garst10,Meng11}. 

Here, we consider quantum phase transitions where the order parameter is characterized by two dynamical exponents $z_>$ and $z_<$, with $z_> > z_<$, and study the scaling behavior of thermodynamics. For renormalization group (RG) fixed-points with multiple $z$, the free energy exhibits only {\it weak} or {\it multiple dynamic scaling}. A similar situation arises in the context of classical critical phenomena \cite{Halperin69,Hohenberg77,Das01,Folk06}, for example, for model C dynamics within the Hohenberg-Halperin classification \cite{Hohenberg77} where the order parameter couples to a conserved density field. 
The dynamical exponents together with the spatial dimension $d$ give rise to two effective dimensions, $D_> = d + z_>$ and $D_< = d+z_<$. Dimensional analysis then requires that the critical part of the free energy density, $\mathcal{F}_{\rm cr}$, is necessarily characterized by two scaling functions, $f_>$ and $f_<$. 
The simplest generalization of the 
scaling hypothesis for $\mathcal{F}_{\rm cr}$ in the presence of multiple dynamic scaling is then given by (in the absence of dangerously irrelevant operators)
\begin{align} \label{SH}
\mathcal{F}_{\rm cr}(r,T) &= \mathcal{F}_{>}(r,T) + \mathcal{F}_{<}(r,T)\nn\\
&= b_1^{-(d+z_>)} f_>(r b_1^{1/\nu}, t_<  b_1^{z_<}, t_>  b_1^{z_>}) \\
&\quad+ b_2^{-(d+z_<)} f_<(r b_2^{1/\nu}, t_<  b_2^{z_<}, t_>  b_2^{z_>})\nn
\end{align} 
with arbitrary scaling parameters $b_1$ and $b_2$. The scaling functions, $f_>$ and $f_<$, depend on the tuning parameter $r$, that measures  the distance to the quantum critical point and scales with the correlation length exponent $\nu$, and, generally, on two temperature scaling fields $t_i = \eta_i T$, with $i \in \{>,<\}$, where $T$ is the temperature and $\eta_i$ are ''kinetic coefficients".
There are three characteristic length scales, $\xi \sim b_{1,2}$, where each of the three arguments of the scaling functions reaches a value of order one, 
\begin{align} \label{scales}
\xi_r \sim  |r|^{-\nu}, \quad
\xi_T^> \sim T^{-1/z_>}, \quad
\xi_T^< \sim T^{-1/z_<}.  
\end{align}
The first can be identified with the correlation length at zero temperature, and the remaining are the two thermal lengths with $\xi_T^> < \xi_T^<$. 

In the limit of small temperatures, $T\to 0$, but finite $r$, we can choose the scaling parameters such that $b_1 = b_2 = |r|^{-\nu} \sim \xi_r$ and the free energy simplifies to
\begin{align}
\mathcal{F}_{\rm cr}(r,T) = |r|^{\nu(d+z_>)} f_>(\sigma, t_> |r|^{-\nu z_>}, t_<  |r|^{-\nu z_<}) 
\nn\\
+ |r|^{\nu(d+z_<)} f_<(\sigma, t_> |r|^{-\nu z_>}, t_<  |r|^{-\nu z_<})
\end{align} 
with $\sigma =$ sign $r$. At $T=0$, the most singular part is then attributed to the contribution with the smaller dynamical exponent $z_<$,
\begin{align} \label{CrLowT}
\mathcal{F}_{\rm cr}(r,0) &\simeq |r|^{\nu(d+z_<)} f_<(\sigma, 0,0)\,.
\end{align} 
So we arrive at the conclusion that the susceptibility $\chi_{rr} = - \partial^2_r \mathcal{F}_{\rm cr}$, that determines, e.g., the correction to the compressibility for pressure tuned quantum critical points, is at $T=0$ determined by the mode with the smaller effective dimension $d+z_<$.

At finite temperatures $T>0$, the generalized scaling hypothesis \eqref{SH}, however, is only of limited use in predicting the critical behavior, in particular, in the regimes $\xi_r > \xi_T^i$. Consider the free energy at scales of the two respective thermal lengths, $b_1 = t_>^{-1/z_>} \sim  \xi_T^>$ and $b_2 = t_<^{-1/z_<} \sim \xi_T^<$,
\begin{align} \label{FT}
\lefteqn{\mathcal{F}_{\rm cr}(r,T) =}
\\\nn 
&= (t_>)^{\frac{d+z_>}{z_>}} f_>\left(r\, t_>^{-\frac{1}{\nu z_>}},  t_< (t_>)^{-\frac{z_<}{z_>}}, 1\right) 
\\\nn 
&+ (t_<)^{\frac{d+z_<}{z_<}} f_<\left(r\, (t_<)^{-\frac{1}{\nu z_<}}, 1, t_>  (t_<)^{- \frac{z_>}{z_<}}\right)\,.
\end{align} 
The temperature dependence is here not explicit due to the residual dependence of the two functions $f_i$ on temperature, so that the critical behavior is not evident. In particular, the third argument of the $f_<$ function diverges for $T \to 0$, and the critical behavior ($r=0$) of thermodynamics will crucially depend on the analytic properties of $f_<$.  

We can distinguish two cases. In the case of {\it decoupled multiple dynamic scaling} the asymptotic behavior of the two functions $f_<$ and $f_>$ in Eq.~\eqref{SH} does not dependent on the respective other thermal fields, $t_>$ and $t_<$,  so that 
the temperature dependence becomes manifest and scaling laws follow with exponents involving either $z_>$ or $z_<$ depending on the quantity of interest. In the case of  {\it coupled multiple dynamic scaling}, however, the dynamics of the two modes mix so that the asymptotic behavior of the two scaling functions does depend on both, $t_>$ and $t_<$, and Eq.~\eqref{FT} is not sufficient to predict the temperature dependence of thermodynamics. In particular, scaling laws can then arise with unusual exponents depending on both dynamical exponents $z_>$ and $z_<$.

In order to identify a mechanism leading to coupled multiple dynamic scaling, we examine a specific example in the following.  We consider the effective theory for a $d$-wave Pomeranchuk (or nematic) instability in two-dimensional isotropic metals describing a spontaneous
deformation of the Fermi surface.
The order parameter is a quadrupolar tensor with two polarizations, $\vec{\Phi}^T = (\phi_1, \phi_2)$, effectively described by a $\Phi^4$-theory
\begin{equation} \label{model}
\mathcal{S} =\int d^dx\,d\tau\,\left[\frac{1}{2}\vec{\Phi}^T \,\mathcal{G}_0^{-1}\,\vec{\Phi} + \frac{u}{4!}\left(\vec{\Phi}^T\,\vec{\Phi}\right)^2\right].
\end{equation}
The dynamics is generated by the coupling of the collective quadrupolar mode to particle-hole pairs in the metal, and, interestingly, it differs for the two polarization of the order parameter. The quadrupolar fluctuations at a given momentum $\bf k$ are damped by particle-hole pairs in the metal if the polarization is longitudinal to  $k_i k_j-\delta_{ij} k^2/2 $ leading to a dynamic exponent $z_> = 3$ while it remains undamped for the transverse polarization giving $z_< = 2$ \cite{Oganesyan01}.
While the renormalization group flow of this model is towards a stable Gaussian fixed-point, it was demonstrated in Ref.~[\onlinecite{Zacharias09}] that the multiple dynamics results in an extended quantum-to-classical crossover. Moreover, it was observed that this extended crossover leads to logarithmic corrections to thermodynamics reminiscent of the coupled multiple scaling discussed above. In order to study the coupling of dynamics in the presence of an interacting fixed-point we generalize the model to an effective dimension $D = 4-\epsilon$ which stabilizes the RG flow towards a Wilson-Fisher (WF) fixed point \cite{Wilson72} so that universal physics at lowest energy emerges and the properties of the scaling functions can be studied.  

The Green function in \eqref{model} possesses the following momentum and frequency dependence \cite{Zacharias09}
\begin{align}
\mathcal{G}_0^{-1}(\vec{q},\omega_n) &= U_{\hat{q}}^{-1} \begin{pmatrix} g^{-1}_<(\vec{q},\omega_n)&0 \\0 & g^{-1}_>(\vec{q},\omega_n)\end{pmatrix}U_{\hat{q}}
\text{ ,}
\end{align}
where the eigenvectors itself depend on the orientation of momentum $\hat q$ via the rotation matrix 
\begin{equation}
U_{\hat{q}} = \begin{pmatrix} \cos\left(2\theta\right)&\sin\left(2\theta\right) \\-\sin\left(2\theta\right) & \cos\left(2\theta\right)\end{pmatrix} ,
\end{equation}
that depends on the angle between momentum and, say, the $x$-axis, $\theta \angle (\vec q, \vec x)$, and reflects the $d$-wave symmetry of the order parameter. In the disordered phase, $\langle \vec \Phi \rangle = 0$, the respective Green functions are given by 
\begin{align}\label{gg}
g^{-1}_<(\vec{q},\omega_n) &= r+q^2+ \eta^2_{<}
\frac{\omega_n^2}{q^{2 z^0_<-2}} \text{ ,}\\
g^{-1}_>(\vec{q},\omega_n) &= r+q^2+\eta_{>}\frac{|\omega_n|}{q^{z^0_>-2}} \text{ ,}\nonumber
\end{align}
with $q = |\vec q|$. The tuning parameter of the quantum phase transition is $r$, and $\eta_{i}$, with $i \in \{>,<\}$, are kinetic coefficients. We consider effective dimensions $d+ z^0_> > 4$ and $d+z^0_< = 4- \epsilon$ so that the whole theory is below its upper critical dimension, $D^+_c = 4$, due to the $z^0_<$-mode. 
We will use $z^0_> = 3$ and discuss the two realizations $d=2$, $z^0_< = 2 - \epsilon$ and $d=2- \epsilon$, $z^0_< = 2$.
The original model is recovered in the limit $\epsilon \to 0$.

\begin{figure}
\centering
$(a)$\includegraphics[scale=.7]{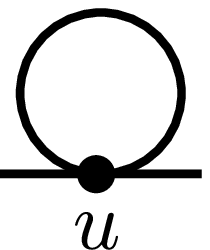}
\quad\quad
$(b)$\quad
\includegraphics[scale=0.7]{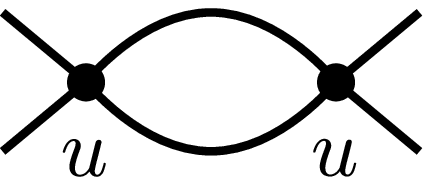}
\caption{One-loop corrections to $(a)$ the tuning parameter and $(b)$ the interaction $u$.}
\label{fig:1loop}
\end{figure} 

We perform a one-loop RG analysis by subsequently integrating out high energy modes within a momentum shell
$(\Lambda/b, \Lambda)$ with $\log b > 1$ and $\Lambda$ being a momentum-cutoff that we set to unity in the following, $\Lambda = 1$.
The form of the theory is restored by rescaling momenta $\vec q\to \vec q/b$ and frequencies $\omega_n \to \omega_n/b^z$. 
To ensure an unbiased treatment of the multiple dynamics, we use here an unspecified dynamical exponent $z$. We will later demonstrate that physical observables do not depend on the specific choice of $z$. As a result of this scaling scheme, the kinetic coefficients, $\eta_i$, obey in one-loop order the tree-level scaling equations 
$\frac{\partial \eta_i}{\partial \log b}  = (z^0_i - z)\,\eta_i$,
where $i \in \{>,<\}$. The quantum fluctuations, see Fig.~\ref{fig:1loop}, lead to an RG flow of the coupling $u$ and the tuning parameter $r$, which at temperature $T=0$ is given by 
\begin{subequations}
\begin{align}
\frac{\partial r}{\partial \log b} &= 2\,r - \frac{1}{24 \pi\eta_<} u\, r \, ,\\
\frac{\partial u}{\partial \log b} &= (4-d-z)\,u - \frac{3}{32 \pi \eta_<} u^2 \,.\label{eq:rg_u}
\end{align}
\end{subequations}
The appearance of the kinetic coefficient $\eta_<$ indicates that the RG flow at $T=0$ is determined by the loop corrections attributed to the $z^0_<$-mode that possesses the smaller effective dimension $d+z^0_< = 4 - \epsilon$, while the $z_>$ mode gives only subleading corrections. Introducing the coupling $U = u/\eta_<$, one finds that for scales  $b > b_{\rm WF} = |U_{\rm WF}/U_0-1|^{1/\epsilon}$ with the bare coupling $U_0$
the interaction is driven towards the WF fixed-point $U \to U_{\rm WF} =  32 \pi \epsilon/3$. Furthermore, from the scaling equation for the tuning parameter follows the correlation length exponent 
\begin{align}
\frac{1}{\nu} = 2 - \frac{4}{9}\epsilon\,.
\end{align}
Note that this value for the correlation length exponent does not belong to the standard universality class of $O(N)$ models with diagonal propagators \cite{Wilson72}. In the following, we consider the theory at scales $b \gg b_{\rm WF}$ where universality emerges and study the RG flow in the vicinity of the WF fixed-point. 

\begin{figure}
\centering
\includegraphics[scale=0.4]{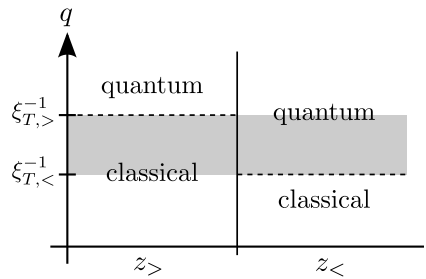}
\caption{Quantum-to-classical crossovers of the $z_>$-mode and $z_<$-mode as a function of momentum. There exists an extended quantum-to-classical crossover regime for $\xi_{T,>}^{-1} > q > \xi_{T,<}^{-1}$ where quantum fluctuations of the $z_<$-mode coexist with thermal fluctuations of the $z_>$-mode (gray shaded area).}
\label{fig:eq2c}
\end{figure}

Temperature $T$ is a relevant scaling field with the scaling dimension given by $z$. For our purposes it is actually convenient to consider the scaling of the two temperature fields $t_> = \eta_> T$ and $t_< = \eta_< T$, 
 \begin{align}
 \frac{\partial t_>}{\partial \log b} = z_>\, t_>\,,\label{eq:t_<}
 \quad {\rm and} \quad 
 \frac{\partial t_<}{\partial \log b} = z_<\, t_<\,,
 \end{align}
 with the identification $z_i = z^0_i$ at one-loop order. The scale where either of the two fields reaches the cutoff scale, $t_i(b_T^i) = 1$,
$i \in \{<,>\}$, defines the two thermal lengths, $b_T^i = \xi_T^i$,
introduced in Eq.~\eqref{scales}. 
At the thermal length scale, the so-called quantum-to-classical crossover \cite{SachdevBook} occurs where critical modes change their character. Whereas for small scales, i.e., for large momenta $q > \xi^{-1}_T$, the critical modes have an essential  quantum character with effective dimension $d+z$, for scales $q^{-1} > \xi_T$ they become classical with effective dimension $d$ only. Due to the presence of multiple thermal lengths, however, the quantum-to-classical crossover in the present case is extended \cite{Zacharias09}. For scales $b$ between the two thermal lengths, $b_T^> < b < b_T^<$, the quantum fluctuations of the $z_<$-mode coexist with the thermal fluctuations of the $z_>$-mode, see Fig.~\ref{fig:eq2c}.
Interestingly, both quantum and classical fluctuations contribute to the RG flow within this range, see Fig.~\ref{fig:1loop}, so that for $b > b_T^> \gg b_{\rm WF}$
\begin{align} \label{ExtXover}
\frac{\partial r}{\partial \log b} &= 2\,r - \frac{u}{24 \pi\eta_<} \, r  + \frac{u T }{6 \pi}=\frac 1 \nu r + \frac{U_{\rm WF} \, t_<}{6 \pi}.
\end{align}
The quantum fluctuations of the $z_<$-mode determine $\nu$ as before. The {\it classical} fluctuations of the $z_>$-mode lead to an additional term linear in $T$. 
This term is crucial as it induces a new length scale that reflects the interplay of the different dynamics. 
The coexistence of quantum and classical fluctuations mixes the scaling fields $r$ and $t_<$ giving rise to a new scaling field  $R(b) = r(b)  +  \frac{\nu U_{\rm WF}}{6 \pi (1 - z_< \nu) } \, t_<(b) $ which obeys
\begin{align} \label{RgR}
\frac{\partial R}{\partial \log b} = \frac{1}{\nu}\,R,
\end{align}
The condition  $R(\xi_R) = 1$ allows to identify the new length scale, $\xi_R =(R(b_T^>))^{-\nu}\,b_T^>$, that reads explicitly 
\begin{align} \label{XiR}
\xi_R =\left(r + \alpha\, T^ {\frac{1}{\nu z_e}} \right)^{-\nu}
\end{align}
with $\alpha=\frac{16 \epsilon}{9} \frac{\nu \, \eta_<}{1- z_< \nu }\,\eta_>^{(1-\nu z_<)/(\nu z_>)} $.
This defines a new `emerging' dynamical exponent given by 
\begin{align} \label{emz}
z_e = \frac{z_>}{\nu (z_> - z_<) + 1}\,.
\end{align}
This simple formula encodes in a nutshell how the interplay of the two time scales can lead to qualitatively new physics. 
At first glance it is surprising that all explicit dependence on the number of space dimensions cancels despite the fact that the contribution arises from {\it classical} fluctuations. The information on the number of dimensions is encoded in the RG equation for $U$. While $u\, T$ has the engineering dimension $4-d$, close to the Wilson-Fisher fixed-point it gets replaced by $U_{\rm WF} t_< $ with dimension $z_<$. For the same reason, there is also no explicit dependence on the interaction strength despite the fact that $d+z_>>4$.

\begin{table}[t]
\centering
\begin{tabular}{|c|c|c|}
\hline
& $d = 2$, $z^0_< = 2-\epsilon$ & $d = 2- \epsilon$, $z^0_< = 2$ 
\\\hline\hline
$z_e$ & $2-\frac{22}{27}\epsilon$ & $2-\frac{4}{27} \epsilon$  
\\[.2em]\hline
$z_e - z_<$ &$\frac{5}{27} \epsilon$& $-\frac{4}{27}\epsilon$  
\\[.2em]\hline\hline
$\chi_{TT}\sim$ & $T^{d/z_>-1} = T^{-1/3}$ & $T^{d/z_>-1} =T^{-(1+\epsilon)/3}$  
\\[.2em]\hline
$\chi_{rT}\sim$ &$T^{(d-1/\nu+z_<)/z_e-1} = T^{7 \epsilon/54}$   & $T^{(d-1/\nu)/z_<}= T^{-5\epsilon/18}$
\\[.2em]\hline
$\chi_{rr}\sim$ & $T^{(d+z_<-2/\nu)/z_e} = T^{-\epsilon/18}$ & $T^{(d+z_<-2/\nu)/z_<} = T^{-\epsilon/18}$  
\\[.2em]\hline
\end{tabular}
\caption{Emergent dynamical exponent $z_e$, see Eq.~\eqref{emz}, for two choices of $d$ and $z^0_<$, and the generalized susceptibilities at criticality $r=0$ as a function of temperature $T$. Coupled multiple dynamic scaling obtains for $z_e > z_<$ with scaling laws depending on $z_e$.
}
\label{table}
\end{table}

The generated temperature dependence in Eq.~\eqref{XiR} results, in principle, in the presence of three length scales, $\xi_T^> \sim T^{-1/z_>}, \xi_T^< \sim T^{-1/z_<}$ and $\xi_R$, even at criticality $r=0$. The generated length $\xi_R \sim T^{-1/z_e}$ obeys $\xi_R/\xi_T^> \gg 1$ as always $z_e < z_>$ but the ratio $\xi_R/\xi^<_T$ depends on the relation between the emerging exponent $z_e$ and $z_<$. Its asymptotic behavior distinguishes whether the multiple dynamic scaling turns out to be coupled or decoupled. 
This is best seen by considering the free energy density of Eq.~\eqref{SH}. For $z_e<z_<$ the RG flow at $r=0$ is cut off by temperature, $T$, before $\xi_R$ is reached. As a result the free energies $\mathcal{F}_i \sim (\xi_T^i)^{-({d+z_i})}$, $i\in \{<,>\}$ are determined by their respective thermal length, $\xi_T^<$ and $\xi_T^>$, so that the multiple dynamic scaling is decoupled, see appendix  for details.
In contrast, for 
\begin{equation}\label{exponents}
z_e>z_< \ \Leftrightarrow \ 1>\nu z_< \ \Leftrightarrow \ 1>\nu z_e
\end{equation}
realized for dimensions $d>2-\frac 5 9 \epsilon$, the two length scales  $\xi_R$ and $\xi_T^>$ govern the critical properties so that $\mathcal{F}_>\sim (\xi_T^>)^{-(d+z_>)}$ and $\mathcal{F}_<\sim (\xi_R)^{-(d+z_<)}$. In particular, the part $\mathcal{F}_<$ is determined by $\xi_R$ and thus depends on the emergent dynamical exponent $z_e$ resulting in coupled multiple dynamic scaling.
Table \ref{table} collects the results for the generalized susceptibilities $\chi_{ab} = - \partial_{a}\partial_b \mathcal{F}_{\rm cr}$ with $a,b = T,r$ at criticality $r=0$. For pressure-tuned quantum criticality, $\chi_{rT}$ and $\chi_{rr}$ can be identified with the thermal expansion $\alpha$ and the critical part of the compressibility, $\kappa$, respectively, and $\chi_{TT}$ is just the specific heat coefficient, $\gamma$. If $z_e > z_<$, we obtain coupled multiple dynamic scaling (i.e.~scaling exponents which depend on both $z_<$ and $z_>$) for the thermal expansion, $\chi_{rT}$, and the compressibility, $\chi_{rr}$.  

Comparing the three thermal length scales with the correlation length at $T=0$, $\xi_r \sim |r|^{-\nu}$, identifies various crossover scales in the phase diagram for $r>0$, see Fig.~\ref{fig:PD}, where the thermodynamics changes its behavior (for more details, see appendix). For coupled multiple dynamic scaling, one crossover line depends on $z_e$.
As usual, for a sufficiently negative $r$ an additional crossover occurs from the quantum WF fixed-point to a classical one giving rise to the Ginzburg line in Fig.~\eqref{fig:PD}. As the model \eqref{model} reduces in the classical limit to a XY-model the classical phase transition is of Kosterlitz-Thouless type for $d=2$.

\begin{figure}
\centering
$(a)$\includegraphics[scale=0.35]{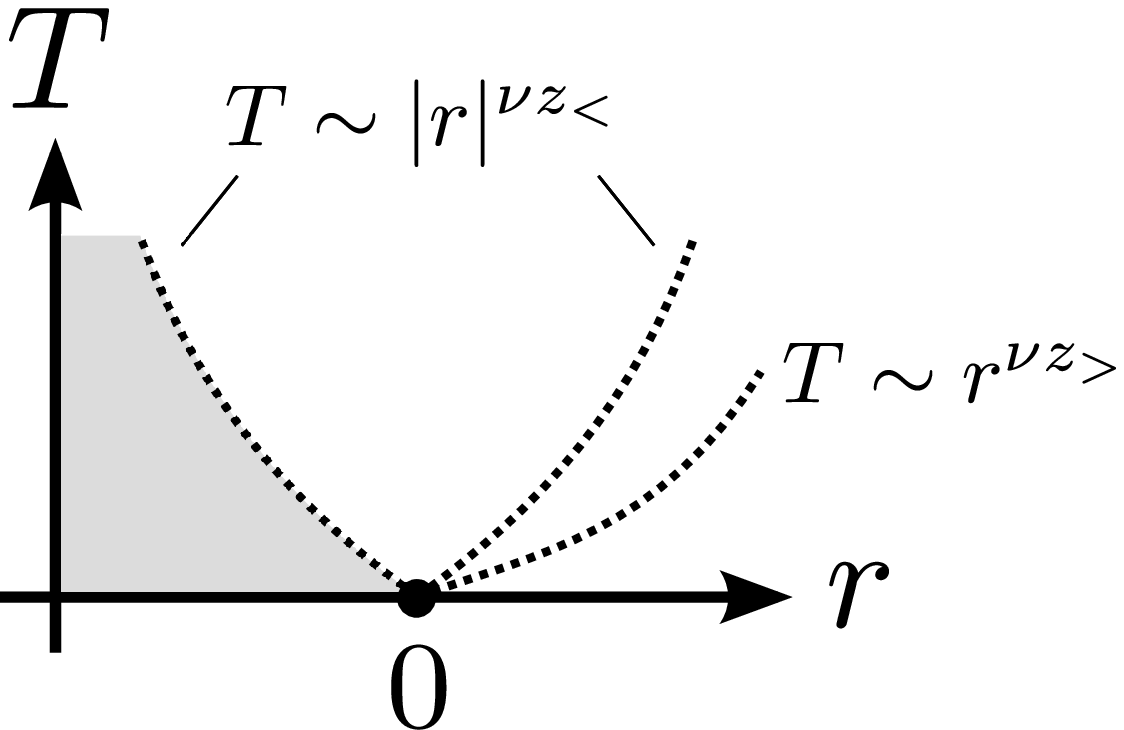}~$(b)$\includegraphics[scale=0.35]{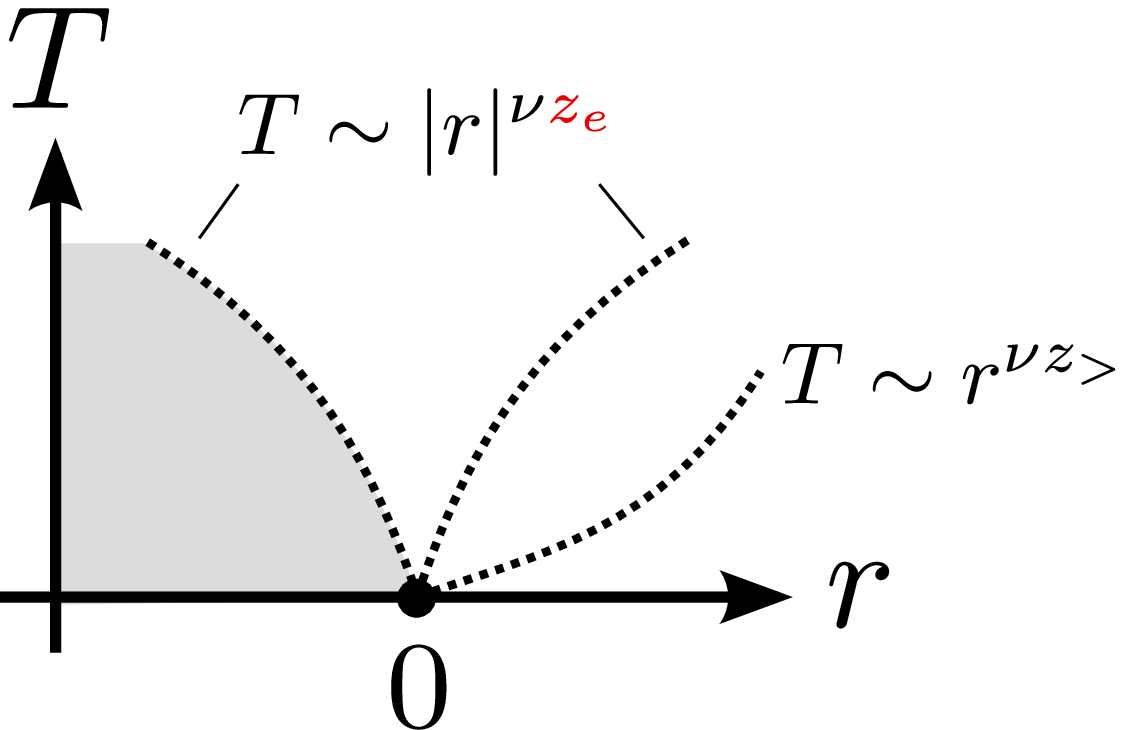}
\caption{Quantum critical phase diagram  and crossover lines (dotted) in the $(r,T)$ plane for
$(a)$ decoupled and $(b)$ coupled multiple dynamic scaling. The shaded area is bounded by the Ginzburg line.
While the specific heat is sensitive only to the lower and the compressibility to the upper crossover line, the
thermal expansion changes its behavior at both crossovers (see appendix).
}
\label{fig:PD}
\end{figure}

To summarize, we studied quantum critical thermodynamics in the presence of multiple dynamical scales, and we distinguished between decoupled and coupled multiple dynamic scaling scenarios. Coupled scaling occurs when  the {\it classical} fluctuations arising
from the mode with the larger dynamical critical exponent generates the mass term which then grows rapidly due to {\it quantum} fluctuations of the other mode.
This quantum `boost' leads  for $\nu z_<<1$ to an emerging thermal length $T^{-1/z_e}$ with the exponent $z_e$ of Eq.~\eqref{emz}. 
Note that this exponent and the mechanism how it is generated are -- at least in one-loop order -- completely universal. $z_e$ depends only on the two dynamical critical exponents $z_<$ and $z_>$ and on the correlation length exponent $\nu$ but not, for example, on the number of space dimensions. As it is also independent of the specific choice of propagators of Eq.~\eqref{gg}, we believe that it
is relevant for a wide class of models.
The additional length scale $T^{-1/z_e}$ gives rise to unusual scaling-laws in thermodynamics. To search for it experimentally, one
option is to look for phase diagrams as shown in Fig.~\ref{fig:PD}$(b)$ where an upper crossover line is convex ($\nu z_e<1$) while a second crossover line has a larger exponent.

\acknowledgements
We acknowledge helpful discussions with M.~Zacharias.  
This work was supported by the DFG through FOR960 and the Bonn-Cologne graduate school (BCGS).

\appendix*
\section{}
The free energy density $\mathcal{F}$ per volume $V$ is defined as usual via the temperature $T$ and the partition function $\mathcal{Z}$ as $ \mathcal{F} = -(T/V) \ln \mathcal{Z}$. The partition function is determined by the functional integral $\mathcal{Z} = \int \mathcal{D}\Phi\, \exp(-\mathcal{S})$ in terms of the action $\mathcal{S}$ defined in Eq.~(6) of the main text.

Introducing a momentum cutoff $|\vec{q}|\leq \Lambda$, we can derive an RG trajectory for the free energy similarly to Refs.~[\onlinecite{nelson75, millis93}]. Integrating out a high energy shell $|\vec{q}|\in[\Lambda/b,\Lambda]$ with $\ln(b) \ll 1$ yields a contribution to the free energy density
\begin{subequations}
\begin{align}
 \delta \mathcal{F} &= \frac{1}{2}\frac{1}{\beta}\sum_{\omega_n}\int_{|\vec{q}|\in[\Lambda/b,\Lambda]} d^dq\left[\ln\left(\beta(r+q^2+\frac{(\eta_<\,\omega_n)^2}{q^{2z_<^0-2}})\right)\right.\nonumber\\
&\qquad\qquad\qquad\qquad+\left.\ln\left(\beta(r+q^2+\frac{|\eta_>\,\omega_n|}{q^{z_>^0-2}})\right)\right]\\
&=  \frac{\Omega_d\,\Lambda^{d}}{(2\pi)^d} \frac{b-1}{b}\,\left[ T \ln\left(2\sinh\left(\frac{\sqrt{\Lambda^{2z_<^0-2}(r+\Lambda^2)}}{2 \eta_<\,T}\right)\right)\nonumber\right.\\
&+ \left.\int_0^{\infty}\frac{d\omega}{2\pi}\coth\left(\frac{\omega}{2T}\right)\arctan\left(\frac{\eta_>\,\omega}{\Lambda^{z_>^0-2}(r+\Lambda^2)}\right) \right]\text{ ,} 
\end{align}
\end{subequations}
with $\Omega_d$ being the $d$-dimensional solid angle. A differential RG equation for $\mathcal{F}$ follows by considering the limit $b\to 1^+$. Furthermore, we note the free energy density has the engineering dimension $d+z$, with $z$ being the arbitrary dynamical exponent used for the scaling of the temperature, $\partial_{\ln(b)}\,T = z\,T$, and thus $T(b) = T\,b^z$. Solving the RG equation one obtains the total free energy density in the form of an integral over an RG trajectory. In one loop order, the latter reads
\begin{equation}
 \mathcal{F} = \int_1^{\infty}db\,b^{-1}\frac{d\mathcal{F}}{d\ln(b)} = \mathcal{F}_> + \mathcal{F}_<\text{ .}
\end{equation}
where

\begin{subequations}
\begin{align}
 \mathcal{F}_> &= \int_1^{\infty}db\,b^{-1}\,b^{-(d+z)}\,\frac{\Lambda^{d}\,\Omega_d}{(2\pi)^d}\,\int_0^{\infty}\frac{d\omega}{2\pi} \coth\left(\frac{\omega}{2T(b)}\right)\nonumber\\
&\quad\times\arctan\left(\frac{\eta_>(b)\,\omega}{\Lambda^{z_>-2}(r(b)+\Lambda^2)}\right)\text{ ,}\\[0.5cm]
 \mathcal{F}_< &= \int_1^{\infty}db\,b^{-1}\,b^{-(d+z)}\,\frac{\Lambda^{d}\,\Omega_d}{(2\pi)^d}\,T(b)\nonumber\\
&\quad\times \ln\left(2\sinh\left(\frac{\sqrt{\Lambda^{2z_<-2}(r(b)+\Lambda^2)}}{2 \eta_<(b)\,T(b)}\right)\right)\text{ .}
\end{align}\label{eq:rg_traject0}
\end{subequations}
As discussed in the main text, the free energy density $\mathcal{F}$ has two contributions stemming from the two modes. 
The scale dependence of the temperature $T(b)$, the kinetic coefficients $\eta_{<,>}(b)$ and the mass $r(b)$ were defined in the main text. In the following, we denote e.g. by $T(b)$ the running temperature and by $T$ without the explicit $b$ dependence the starting value for the RG flow. 
Redefining first $\omega = \epsilon\,T(b)$ and then substituting $\Lambda/b = t_<^{1/z_<}\,e^{-l}$ for the part $\mathcal{F}_<$ and $\Lambda/b = t_>^{1/z_>}\,e^{-l}$ for $\mathcal{F}_>$, with $t_i = \eta_i\,T$, we obtain
\begin{subequations}
\begin{align}
 \mathcal{F}_> &= \frac{t_>^{(d+z_>)/z_>}}{\eta_>}\int_0^{\infty}dl\,e^{-dl}\,\frac{\Omega_d}{(2\pi)^d}\,\int_0^{\infty}\frac{d\epsilon}{2\pi}\coth\left(\frac{\epsilon}{2}\right)\nonumber\\
&\quad\times\arctan\left(\frac{\epsilon\,e^{z_>l}}{r(\Lambda\,t_>^{-1/z_>}\,e^{l})/\Lambda^2+1}\right)\text{ ,}\label{eq:f_larger}\\[0.5cm]
 \mathcal{F}_< &= \frac{t_<^{(d+z_<)/z_<^0}}{\eta_<}\int_0^{\infty}dl\,e^{-dl}\,\frac{\Omega_d}{(2\pi)^d}\label{eq:f_smaller}\\
&\times \ln\left(2\sinh\left(\frac{\sqrt{r(\Lambda\,t_<^{-1/z_<}\,e^{l})/\Lambda^{2}+1}}{2\,e^{z_<l}}\right)\right).
\nonumber
\end{align}\label{eq:rg_traject}
\end{subequations}
Note that with these substitutions all dependencies on the arbitrarily chosen scaling exponent $z$ have vanished. 
Next, we write Eqs.~\eqref{eq:rg_traject} in a scaling form. For $T>r^{\nu z_>}$, the flowing mass can be expressed in terms of the emergent scaling field $R(b)$ and the constant $\tilde{\alpha} = (\nu U_{\rm WF})/(6\pi(1-z_<\nu))$,

\begin{align}
 r(b) &= R(b)-\tilde{\alpha}\,\eta_<(b)\,T(b)\text{ .}
\end{align}
With $R(b) = R \,b^{1/\nu}=\Lambda^2\,\xi_R^{-1/\nu}\,(\Lambda/b)^{-1/\nu}$ and $\eta_<(b)\,T(b) = t_<\,b^{z_<}$ (see Eqs.~\eqref{eq:t_<} and \eqref{RgR} of the main text) we find that the free energy density $\mathcal{F}_<$ depends on

\begin{align}
 &\Lambda^{-2}r(\Lambda\,t_<^{-1/z_<}\,e^{l}) \\\nonumber&= (t_<^{1/z_<}\,\xi_R)^{-1/\nu}\,e^{l/\nu}-\tilde{\alpha}\,\Lambda^{2_<-2}\,e^{z_<l}\text{ ,}
\end{align}
and therefore only on the combination $t_<^{-1/(\nu z_<)}\,R$. It can thus be expressed in the scaling form

\begin{align}
 \mathcal{F}_< = b^{-(d+z_<)}\,\frac{1}{\eta_<}\,\hat{f}_<(R\,b^{1/\nu},t_<\,b^{z_<})\text{ .}
\end{align}
From the usual scaling analysis one can now deduce that the free energy density of the mode with the smaller dynamical exponent scales for $T>r^{\nu z_>}$ as

\begin{align}
  \mathcal{F}_< \sim \begin{cases}(\xi_R)^{-(d+z_<)}&\text{if}~~z_e>z_<~\Leftrightarrow~\xi_R\ll\xi_T^<\\(\xi_T^<)^{-(d+z_<)}&\text{if}~~z_e<z_<~\Leftrightarrow~\xi_R\gg\xi_T^<\end{cases}\text{ .}
\end{align}
This result can also be obtained explicitly by performing the integral in Eq.~\eqref{eq:f_smaller}. In a similar manner, we find that the free energy density $\mathcal{F}_>$ can be written in terms of the scaling function
\begin{align}
 \mathcal{F}_> = b^{-(d+z_>)}\,\frac{1}{\eta_>}\,\hat{f}_>(R\,b^{1/\nu},t_<\,b^{z_<},t_>\,b^{z_>})
\end{align}
in the regime $T>r^{\nu z_>}$. Since however $\xi_T^> \ll \xi_R, \xi_T^<$ in this regime, we find that the free energy density of the mode with the larger dynamical exponent always scales for $T>r^{\nu z_>}$ as

\begin{align}
 \mathcal{F}_> \sim (\xi_T^>)^{-(d+z_>)}\text{ .}
\end{align}
This can also be derived by explicit integration of the RG trajectory \eqref{eq:f_larger}.

Thermodynamic observables are defined as derivatives of the total free energy density $\mathcal{F} = \mathcal{F}_<+\mathcal{F}_>$ with respect to either the temperature $T$ or the tuning parameter $r$. 
Here, $T$ and $r$ are the starting values of the RG trajectories, which are used in Eqs.~\eqref{eq:rg_traject} to determine the free energy.
%
%They can thus be calculated by acting with these derivatives on the RG trajectories \eqref{eq:rg_traject}. 
In the following, we focus on the critical, i.e., the most singular contributions to thermodynamics close to the quantum critical point. The expression of the flowing mass $r(b)$ appearing in the RG trajectories for the free energy densities can be obtained by integrating the corresponding RG equation \eqref{ExtXover} given in the main text. For RG stages $b \gg b_{\rm WF}\gg1$ with $b_{\rm WF}$ being the RG scale at which the Wilson-Fisher fixed point is reached, we obtain that
\begin{align}
 r(b) &= r\,\left(U_{\rm WF}/U_0\right)^{4/9}\,b^{1/\nu}\\
&+\frac{K\,U_{\rm WF}}{1/\nu-z_<} \eta_{<}\,T\,\left(b^{1/\nu}\,b_T^>{}^{z_<-1/\nu}-b^{z_<}\right)\,\Theta(b-b_T^>)\text{ ,}\nonumber
\end{align}
where $U_0 = \Lambda^{-\epsilon} u_0/\eta_<$ with $\epsilon = 4-d-z_<$ being the bare reduced interaction and $U_{\rm WF}$ its Wilson-Fisher fixed point value, and the constant $K=\Lambda^{d-2}\,\Omega_d/(3(2\pi)^d)$. A derivative with respect to tuning parameter $r$ therefore acts on the running $r(b)$ as

\begin{align}
 \frac{d r(b)}{dr} = \left(U_{\rm WF}/U_0\right)^{4/9}\,b^{1/\nu}\text{ ,}
\end{align}
 and similar terms arise for derivatives with respect to the temperature if the RG flow reaches the extended quantum to classical crossover regime for sufficiently high temperatures $T$.

\begin{figure}
\centering
$(a)$\includegraphics[scale=0.35]{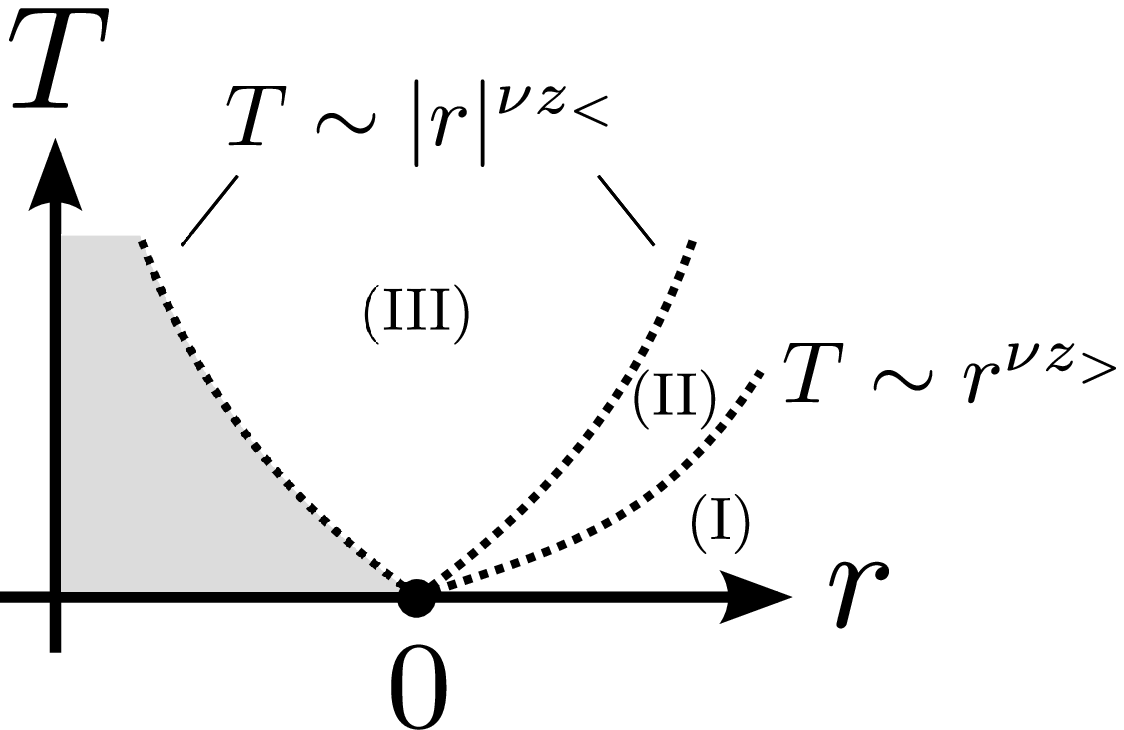}~$(b)$\includegraphics[scale=0.35]{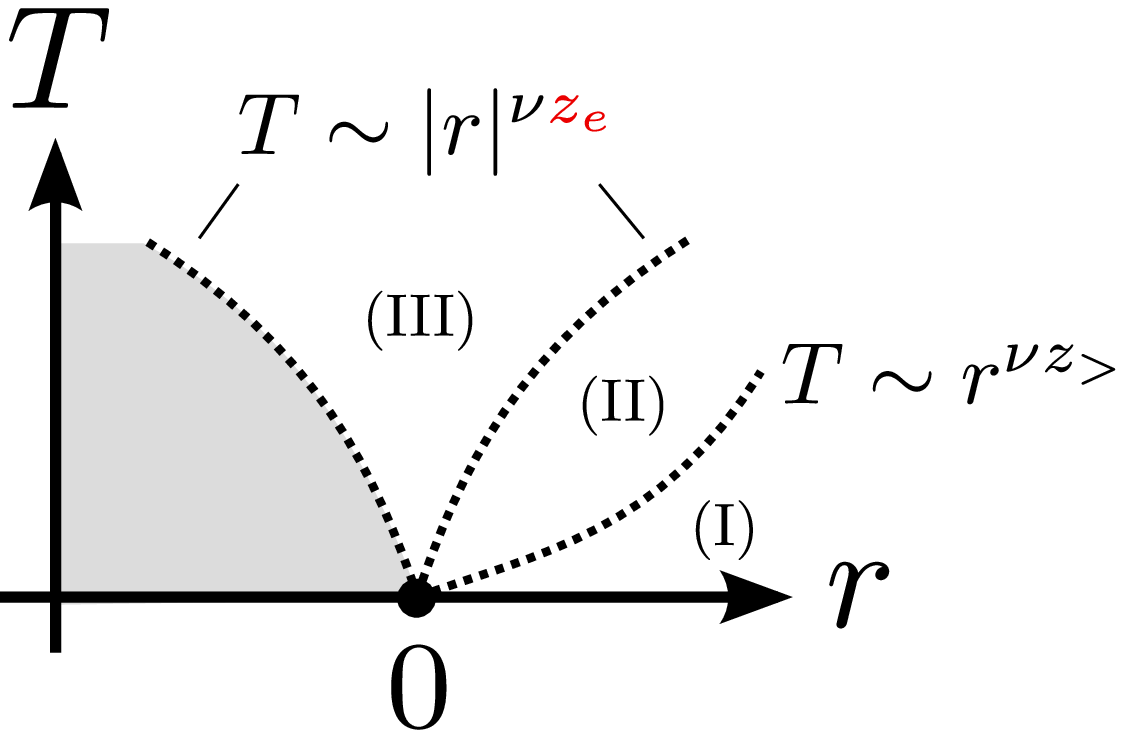}
\caption{Generic phase diagrams for systems with multiple dynamics. Subfigure $(a)$ and $(b)$ shows the cases of \textit{decoupled}, $\nu z_< > 1$, and \textit{coupled}, $\nu z_< < 1$, multiple dynamic scaling, respectively. The regime (II) arises due to the presence of multiple thermal scales.}
\label{fig:PD_append}
\end{figure} 

Each of the two parts, $\mathcal{F}_>$ and $\mathcal{F}_<$, of the free energy density possesses its own crossover between a high and low temperature limit. They are determined by the comparison of the correlation length at $T=0$, $\xi_r = r^{-\nu}$, with the corresponding thermal length. For $\mathcal{F}_>$, the crossover occurs at $T\sim r^{\nu z_>}$, while it is located at $T\sim r^{\nu z_{(\cdot)}}$ with $z_{(\cdot)} = \max\{z_<,z_e\}$ for $\mathcal{F}_<$. These crossovers are directly reflected in the behavior of thermodynamic observables. We find that the critical part of the specific heat, $c = -T\,\partial_T^2 \mathcal{F}$, is only sensitive to the lower crossover line between regime (I) and (II), see Fig.~\ref{fig:PD_append},
\begin{align}
c  \sim \begin{cases}T \,r^{\nu_{}(d-z_>)}  ~&{\rm if}\quad T\ll r^{\nu_{} z_{>}}\,, {\rm (I)}\\
                  T^{d/z_>}~&{\rm if}\quad T\gg r^{\nu_{} z_{>}}, {\rm (II)+(III)}
                 \end{cases}\text{ .}
\end{align}
It is always dominated by the mode with the larger dynamical exponent $z_>$ because the latter has the larger phase space. 

The critical part of the compressibility, $\kappa = -\partial_r^2 \mathcal{F}$, on the other hand, is only sensitive to the upper crossover line between regime (II) and (III),
\begin{align}
\kappa  \sim \begin{cases}r^{\nu_{}(d+z_<)-2}~&{\rm if}\quad T\ll r^{\nu_{} z_{(\cdot)}} \,, {\rm (I)+(II)} \\
T^{(d+z_<-2/\nu_{})/z_{(\cdot)}}~&{\rm if}\quad T\gg r^{\nu_{} z_{(\cdot)}} \,, {\rm (III)}
\end{cases}
\end{align}
with $z_{(\cdot)} = \max\{z_<,z_e\}$. 
The most singular part of the compressibility is attributed to the fluctuations with the smaller dynamical exponent $z_<$.

Most interesting is the thermal expansion $\alpha = \partial_T\partial_r\mathcal{F}$, as it is sensitive to both crossovers. 
We find that the behavior of $\alpha$ at lowest temperatures $T < r^{\nu z_>}$ is dominated by the mode with the larger dynamical exponent $z_>$. At higher temperatures, the scaling depends on whether decoupled or coupled multiple dynamic scaling is obtained. For \textit{decoupled} scaling present for $\nu z_< > 1$, the critical part reads
\begin{align}
 \alpha \sim  \begin{cases}T \,r^{\nu(d-1/\nu_{}-z_>)}  &{\rm if}\quad T\ll r^{\nu_{} z_{>}}  \,, {\rm (I)}\\
                  r^{\nu_{}(d-1/\nu_{})}&{\rm if}\quad  r^{\nu_{} z_{<}} \gg T\gg r^{\nu_{} z_{>}}  \,, {\rm (II)}\\
  T{}^{(d-1/\nu_{})/z_<}&{\rm if}\quad  T\gg r^{\nu_{} z_{<}}  \,, {\rm (III)}.
                 \end{cases} 
\end{align}
For \textit{coupled} scaling, $\nu z_< < 1$, on the other hand, we find that the thermal expansion is sensitive the emergent dynamical exponent. We obtain
\begin{align}
  \alpha \sim  \begin{cases}T \,r^{\nu(d-1/\nu_{}-z_>)}  &{\rm if}\quad T\ll r^{\nu_{} z_{>}}  \,, {\rm (I)}\\
                  T^{1/(\nu z_e)-1}\,r^{\nu(d+z_<)-2}&{\rm if}\quad r^{\nu_{} z_{e}} \gg T\gg r^{\nu_{} z_{>}}  \,, {\rm (II)}\\
  T{}^{(d+z_<-z_e-1/\nu_{})/z_e}&{\rm if}\quad T\gg r^{\nu_{} z_{e}}  \,, {\rm (III)}.
                 \end{cases} 
\end{align}
In order to check that the leading behavior indeed matches at the lower crossover $T \sim r^{\nu z_>}$, one needs the explicit definitions of $\nu$ and $z_e$ as given in Eqs.~\eqref{ExtXover} and \eqref{emz} of the main text. In conclusion, we find that the thermal expansion is sensitive to both crossovers in the phase diagram of Fig.~\ref{fig:PD_append} in agreement with the study of Ref.~\onlinecite{Zacharias09}.  

In the end, we would like to discuss the curvature of the upper crossover line between regime (II) and (III). 
For decoupled and coupled multiple dynamic scaling the exponents fulfill the following relations,
\begin{align}
\begin{array}{ll}
{\rm decoupled:}& \nu\,z_< > 1 \\
{\rm coupled:}&  \nu\,z_< < \nu\,z_e < 1.
\end{array}
\end{align}
As a result, the upper crossover line is concave for coupled while it convex for decoupled scaling. This serves as a convenient necessary condition to identify the type of scaling at work, which can be used in experiments.

%%%%%%%%%%%%%%%%%%%%%%%


\begin{thebibliography}{99}

\bibitem{SachdevBook}
S. Sachdev, {\it Quantum Phase Transitions}, (Cambridge University Press, 2011).

\bibitem{Loehneysen07}
H. v.	L\"ohneysen, A. Rosch, M. Vojta, and P. W\"olfle,
Rev.	Mod. Phys. {\bf 79}, 1015 (2007).

\bibitem{Gegenwart07}
P. Gegenwart {\it et al.},
%Westerkamp, T., Krellner, C., Tokiwa, Y., Paschen, S., Geibel, C., Steglich, F., et al. (2007). Multiple Energy Scales at a Quantum Critical Point. 
Science {\bf 315}, 969 (2007).

\bibitem{Belitz01}
D. Belitz, T. R. Kirkpatrick, M. T. Mercaldo, and S. L. Sessions, Phys. Rev. B {\bf 63}, 174427 (2001).
\bibitem{Metlitski10}
Max A. Metlitski and Subir Sachdev, Phys. Rev. B {\bf 82}, 075127 (2010); {\it ibid.} {\bf 82} 075128 (2010).

\bibitem{Zacharias09}
M. Zacharias, P. W\"olfle, and M. Garst,
Phys. Rev. B {\bf 80}, 165116 (2009).

\bibitem{Garst10}
M. Garst and A. V. Chubukov,
Phys. Rev. B {\bf 81}, 235105 (2010).

\bibitem{Meng11}
T. Meng, M. Dixit, M. Garst, and J. S. Meyer,
Phys. Rev. B, {\bf 83}, 125323 (2011).

\bibitem{Halperin69}
B. I. Halperin and P. Hohenberg, 
Phys. Rev. {\bf 177} 952 (1969).

\bibitem{Hohenberg77}
P. C. Hohenberg and B. I. Halperin,
Rev. Mod. Phys. {\bf 49}, 435 (1977).

\bibitem{Das01}
D. Das, A. Basu, M. Barma, and S. Ramaswamy, 
Phys. Rev. E {\bf 64}, 021402 (2001).

\bibitem{Folk06}
R. Folk and G. Moser,
Journal of Physics A, {\bf 39} R207 (2006).


\bibitem{Oganesyan01}
V. Oganesyan, S. A. Kivelson, and E. Fradkin, 
Phys. Rev. B {\bf 64}, 195109 (2001).

\bibitem{Wilson72}
K. G. Wilson and M. E. Fisher, Phys. Rev. Lett. {\bf 28}, 240Ð243 (1972).

\bibitem{nelson75}
D. R. Nelson, Phys. Rev. B \textbf{11}, 3504 (1975).

\bibitem{millis93}
A. J. Millis, Phys. Rev. B \textbf{48}, 7183 (1993).


\end{thebibliography}
\end{document}